\documentclass[12pt]{article}
\usepackage{amsthm,amsfonts,amsmath,amssymb,mathrsfs,graphicx,url,hyperref}

\title{On a Derivation of the Absorbing Boundary Rule}

\author{
	Roderich Tumulka\footnote{Fachbereich Mathematik, Eberhard-Karls-Universit\"at T\"ubingen, Auf der Morgenstelle 10, 72076 T\"ubingen, Germany. E-mail: roderich.tumulka@uni-tuebingen.de}
}
\date{September 22, 2023}

\newcommand{\be}{\begin{equation}}
	\newcommand{\ee}{\end{equation}}

\renewcommand{\Im}{\mathrm{Im}}

\newcommand{\PPP}{\mathbb{P}}
\newcommand{\RRR}{\mathbb{R}}

\newcommand{\vj}{\boldsymbol{j}}
\newcommand{\vn}{\boldsymbol{n}}

\newcommand{\vy}{\boldsymbol{y}}

\newcommand{\vx}{\boldsymbol{x}}
\newcommand{\vX}{\boldsymbol{X}}
\newcommand{\vzero}{\boldsymbol{0}}

\newcommand{\sZ}{\mathscr{Z}}

\begin{document}
\maketitle
\begin{abstract}
Consider detectors waiting for a quantum particle to arrive at a surface $S$ in 3-space. For predicting the probability distribution of the time and place of detection, a rule was proposed in \cite{Tum16}, called the absorbing boundary rule (ABR) and involving a 1-particle Schr\"odinger equation with an absorbing boundary condition on $S$. While plausibility arguments for the ABR were given there, it would be desirable to derive the ABR from a microscopic model of a detector. We outline here such a derivation by putting together known results from the literature. Our derivation is non-rigorous, and it would still be desirable to have a rigorous version of it in the future.
		
\medskip
		
\end{abstract}

\section{Introduction}
\label{sec:intro}

There has long been interest (see, e.g., \cite{AB61,MSE08,MRC09} and references therein) in the question how to compute the probability distribution of the time when (and place where) detectors sitting along the surface $S=\partial \Omega$ of a region $\Omega \subset \RRR^3$ in physical 3-space will be triggered by a quantum particle with initial wave function $\psi_0\in L^2(\Omega)$. A proposed distribution, intended to be an approximation close to the observed distribution if the actual detectors can be regarded as close approximations to ideal detectors, was formulated as the \emph{absorbing boundary rule} (ABR) in \cite{Tum16}; we repeat it in Section~\ref{sec:ABR} below. The derivation of the ABR given in \cite{Tum16} has the status of plausibility arguments, and it would be of interest to see a derivation (rigorous, if possible) of it from a microscopic model of the detectors as quantum systems. Here, we outline such a derivation in a non-rigorous way; another derivation, based on repeated position measurements on a particle on a lattice and taking a suitable continuum limit, was given by Dubey et al.\ \cite{DBD20}.

The present derivation is based on putting together several steps that can be found in the literature (more detail in Section~\ref{sec:derive}) as follows:

\begin{enumerate}
\item We model the detector as a volume filled with atoms that can interact locally with the test particle (i.e., the quantum particle to be detected), as in a cloud chamber. For this situation, it has been argued (specifically by Ballesteros et al.\ \cite{BBFF2020} but similarly also in other works on continuous measurement \cite{Dio}) that the conditional wave function $\psi$ of the test particle follows a stochastic process approximately given by the Ghirardi-Rimini-Weber (GRW) collapse process \cite{GRW86,Bell87a} or a similar one, where each collapse is associated with a detection of the test particle.

\item We make the detection rate position (and possibly time) dependent to reflect the fact that the detector atoms are confined to a certain volume; this leads to a different form of the stochastic process that was described in \cite{Tum05}.

\item We consider the process only up to the first detection, which leads to an effective time evolution given by a 1-particle Schr\"odinger equation with an imaginary potential. Imaginary potentials were used long before to model particle absorption or detection, specifically by Bethe \cite{Bet40} and Allcock \cite{All69b}.

\item As pointed out in \cite{Tum19}, the absorbing boundary condition arises as a limiting case of imaginary potentials.
\end{enumerate}

In the remainder of this paper, we provide details about these steps. For simplicity, we limit ourselves to the non-relativistic case without spin and assume that the surface $\partial\Omega$ along which the detectors are placed is time independent (for an ABR for moving detectors, see \cite{Tum16b}).

\section{Statement of the Absorbing Boundary Rule}
\label{sec:ABR}

We assume that the initial wave function $\psi_0\in L^2(\Omega)$ (with $\|\psi_0\|=1$) is prepared at time $t=0$. A detection event can occur at a random time $T>0$ and place $\vX\in\partial\Omega$; the outcome $Z$ of the experiment is the pair $(T,\vX)$. If the test particle never gets detected, then we write $Z=\infty$, so the set $\sZ$ of possible outcomes is $\sZ=\bigl([0,\infty)\times\partial\Omega \bigr)\cup\{\infty\}$. The ABR demands that we solve, with initial datum $\psi_0$, the 1-particle Schr\"odinger equation
\be\label{Schr}
i\hbar \frac{\partial\psi}{\partial t} = -\frac{\hbar^2}{2m} \nabla^2\psi+ V\psi
\ee
in $\Omega$ subject to the boundary condition
\be
\vn(\vx)\cdot\nabla \psi(\vx) = i\kappa(\vx)\, \psi(\vx) 
\ee
at every $\vx\in\partial\Omega$, where $\vn(\vx)$ is the outward unit normal vector to $\partial\Omega$ at $\vx$ and $\kappa(\vx)>0$ is a given constant, the detector parameter at $\vx$.
(This evolution is mathematically well defined \cite{TT}.) Then the probability distribution of $Z$ on $\sZ$ is given by
\begin{equation}\label{probnjR}
  \PPP_{\psi_0} \Bigl( t \leq T<t+dt, \vX \in d^2\vx \Bigr) =
  \vn(\vx) \cdot
  \vj^{\psi_t}(\vx) \, dt \, d^2\vx
\end{equation}
with $d^2\vx$ denoting a surface element or its area and $\vj^\psi$ the probability current vector field
defined by $\psi$, i.e.,
\begin{equation}\label{jSchr}
  \vj^\psi = \frac{\hbar}{m} \Im\, \psi^* \nabla \psi\,,
\end{equation}
together with
\be\label{Zinfty}
\PPP_{\psi_0}(Z=\infty) = 1-  \int\limits_{0}^{\infty} dt \int\limits_{\partial\Omega} d^2\vx \; \vn(\vx) \cdot
  \vj^{\psi_t}(\vx)\,.
\ee

\section{Derivation}
\label{sec:derive}

We now go through the derivation, with the subsections numbered according to the steps 1--4 listed in the introduction.

\subsection{Continuous Measurement and the GRW Process}

For a test particle moving through a cloud chamber, it has been argued \cite{BBFF2020} that the wave function of the test particle will effectively collapse according to a 1-particle GRW process $\Psi_t$,\footnote{The process described in \cite{BBFF2020} differs slightly from a GRW process in that the collapse times are not exponentially distributed, but the exponential distribution is actually appropriate here.} which can be described as follows \cite{Bell87a} for a given initial wave function $\Psi_0\in L^2(\RRR^3)$ (with $\|\Psi_0\|=1$) at time $t=0$: The collapse times $0<T_1<T_2<\ldots$ are random and form a Poisson process with rate $\lambda>0$. Between 0 and $T_1$, and between $T_k$ and $T_{k+1}$, $\Psi_t$ evolves unitarily according to the Schr\"odinger equation
\be\label{Schr2}
i\hbar \frac{\partial\Psi}{\partial t} = -\frac{\hbar^2}{2m} \nabla^2\Psi+ V\Psi =: H\Psi
\ee
in $\RRR^3$ (without boundary condition). At $T_k$, $\Psi_t$ jumps according to
\be\label{GRW1}
\Psi_{T_k+}(\vx)= \frac{g(\vx-\vX_k)^{1/2} \,\Psi_{T_k-}(\vx)}{\bigl( \int d^3\vy \: g(\vy-\vX_k) \, |\Psi_{T_k-}(\vy)|^2\bigr)^{1/2}}\,,
\ee
where the notation $f_{t\pm}$ means $\lim_{\tau\searrow0} f_{t\pm \tau}$,
\be
g(\vx)=(2\pi\sigma^2)^{-3/2}\exp(-\vx^2/2\sigma^2)
\ee
is the Gaussian density with mean $\vzero$ and width $\sigma$, and the collapse center $\vX_k$ is chosen randomly with probability distribution
\be\label{GRW2}
\PPP\bigl(\vX_k\in d^3\vx \big| \Psi_{T_k-} \bigr) = d^3\vx \int d^3\vy \: g(\vy-\vx) \, |\Psi_{T_k-}(\vy)|^2 \,.
\ee
Here, $\sigma$ and $\lambda$ are parameters of the process, the collapse width and collapse rate.

\subsection{Position-Dependent Collapse Rate}

The variant of the GRW process for position-dependent collapse rates was described in \cite{Tum05}. Suppose we are given, for every $\vx\in\RRR^3$, a positive operator $\Lambda(\vx)$ called the collapse rate operator. 
A collapse event at $(t,\vx)$ occurs with rate $\langle \Psi_t|\Lambda(\vx)|\Psi_t\rangle$, and in case of an event at $(T,\vX)$, $\Psi$ jumps according to
\be
\Psi_{T+} = \frac{\Lambda(\vX)\, \Psi_{T-}}{\|\Lambda(\vX)\, \Psi_{T-}\|}\,.
\ee
Between collapses, $\Psi_t$ evolves non-unitarily according to
\be\label{semigroup1}
i\hbar\frac{\partial\Psi_t}{\partial t} = \Bigl(H-\tfrac{i\hbar}{2}\int d^3\vx \, \Lambda(\vx)\Bigr) \Psi_t \,.
\ee

The original GRW process as in \eqref{GRW1} and \eqref{GRW2} corresponds to the choice
\be\label{Lambdalambda1}
(\Lambda(\vx) \psi)(\vy) = \lambda \, g(\vx-\vy) \, \psi(\vy)\,,
\ee
and the process with position-dependent collapse rate $\lambda(\vx)$ to
\be\label{Lambdalambda2}
(\Lambda(\vx) \psi)(\vy) = \lambda(\vx) \, g(\vx-\vy) \, \psi(\vy)\,.
\ee
As a consequence, in this case the collapse rate is given by
\be
\langle \Psi_t|\Lambda(\vx)|\Psi_t\rangle = \lambda(\vx) \int d^3\vy \, g(\vx-\vy) \, |\Psi_t(\vy)|^2 = \lambda(\vx) \, (g \ast |\Psi_t|^2)(\vx)\,,
\ee
where $\ast$ means convolution. In particular, on a length scale $\gg \sigma$, convolution with $g$ does not change much, so the collapse rate is approximately given by $\lambda(\vx) \, |\Psi_t(\vx)|^2$.

\subsection{Imaginary Potential}

Since we are only interested in the evolution up to the first detection event, we can say that, on a length scale $\gg\sigma$, detection occurs at $(t,\vx)$ with rate $\lambda(\vx) \, |\Psi_t(\vx)|^2$, where $\lambda(\vx)$ represents the strength of the detector at $\vx$, and $\Psi_t$ evolves according to \eqref{semigroup1}. Since, by \eqref{Lambdalambda2}, all $\Lambda(\vx)$ are multiplication operators, so is $\int d^3\vx\, \Lambda(\vx)$. In fact, it is multiplication by $\lambda \ast g$, which on a length scale $\gg\sigma$ is approximately $\lambda(\vx)$. In this approximation, the Schr\"odinger equation becomes
\be\label{Schr3}
i\hbar \frac{\partial\Psi}{\partial t} = -\frac{\hbar^2}{2m} \nabla^2\Psi(\vx)+ V(\vx)\Psi(\vx)-\tfrac{i\hbar}{2}\lambda(\vx) \Psi(\vx)\,.
\ee
This is exactly how imaginary potentials are used to model soft detectors (cf., e.g., \cite[Rem.~3]{Tum16b}): the additional potential $\tilde V$ is negative-imaginary, and the detection (and absorption) rate at $\vx$ is $(2i/\hbar) \tilde V(\vx) \, |\Psi_t(\vx)|^2$.

\subsection{Absorbing Boundary Condition}

As derived in \cite{Tum19}, the ABR can be obtained from soft detectors modeled through imaginary potentials by considering a thin layer of soft detectors around $\partial\Omega$ together with a reflecting Neumann (or Robin) boundary \cite{Tum23} at the outer surface of the layer. One then takes the limit in which the strength $\lambda(\vx)$ of the soft detector increases to $\infty$ and the thickness $L(\vx)$ of the layer decreases to 0 so that $\lambda(\vx) \, L(\vx)\to \hbar\kappa(\vx)/m$.

This completes our outlined derivation of the absorbing boundary rule.

\end{document}